\documentclass[preprint,12pt]{elsarticle}

\usepackage{lineno}
\usepackage{amssymb}
\usepackage{subcaption}
\usepackage{comment}
\usepackage{amsmath}
\newcommand{\coatjava}{\textsc{coatjava }}

\def\realnumbers{\mathbb{R}}

\usepackage{lineno}

\journal{Nuclear Physics A}

\begin{document}
\begin{frontmatter}



\title{AI-Assisted Object Condensation Clustering for Calorimeter Shower Reconstruction at CLAS12}


\author[aff1]{Gregory Matousek\corref{cor1}}
\cortext[cor1]{Corresponding author. \textit{E-mail address}: gregory.matousek@duke.edu}
\author[aff2]{Anselm Vossen}

\affiliation[aff1]{organization={Duke University},
            addressline={120 Science Drive}, 
            city={Durham},
            postcode={27708}, 
            state={NC},
            country={USA}}

\affiliation[aff2]{organization={Thomas Jefferson National Accelerator Facility},
            addressline={12000 Jefferson Avenue}, 
            city={Newport News},
            postcode={23606}, 
            state={VA},
            country={USA}}

\begin{abstract}
Several nuclear physics studies using the CLAS12 detector rely on the accurate reconstruction of neutrons and photons from its forward angle calorimeter system. These studies often place restrictive cuts when measuring neutral particles due to an overabundance of false clusters created by the existing calorimeter reconstruction software. In this work, we present a new AI approach to clustering CLAS12 calorimeter hits based on the object condensation framework. The model learns a latent representation of the full detector topology using GravNet layers, serving as the positional encoding for an event's calorimeter hits which are processed by a Transformer encoder. This unique structure allows the model to contextualize local and long range information, improving its performance. Evaluated on one million simulated $e^-+p$ collision events, our method significantly improves cluster trustworthiness: the fraction of reliable neutron clusters, increasing from 8.88\% to 30.73\%, and photon clusters, increasing from 51.07\% to 64.73\%. Our study also marks the first application of AI clustering techniques for hodoscopic detectors, showing potential for usage in many other experiments. 
\end{abstract}

\begin{graphicalabstract}
\begin{figure}[htbp]
    \centering
    \includegraphics[width=\textwidth]{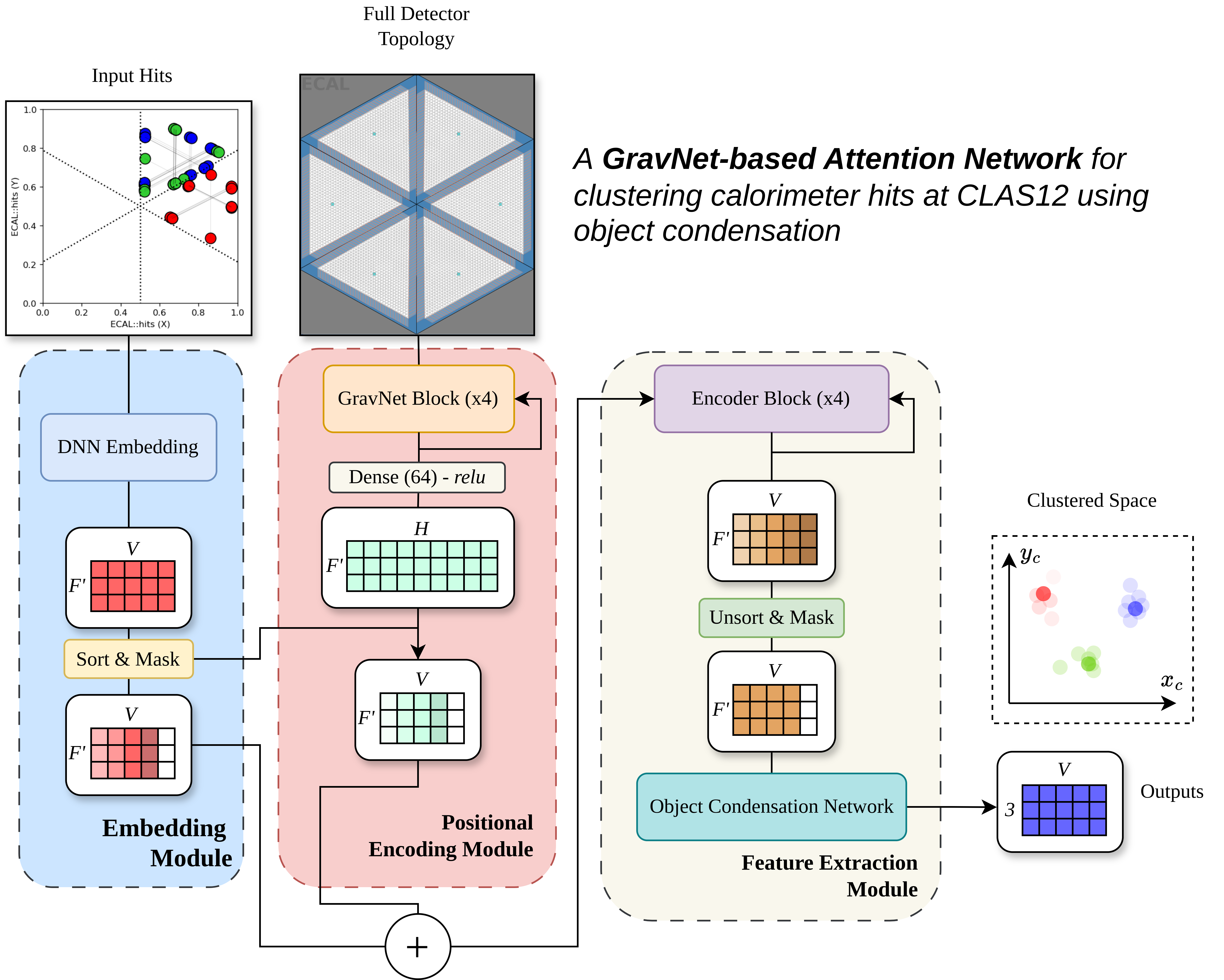}
    \caption*{Schematic of clustering network architecture.}
    \label{fig:graphical_abstract}
\end{figure}
\end{graphicalabstract}

\begin{highlights}
\item Introduces an AI-based clustering method for CLAS12's electromagnetic calorimeter.
\item Uses GravNet and a Transformer encoder to learn hit representations.
\item Implements object condensation as a framework to perform hit clustering.
\item First AI clustering method applied to hodoscopic detectors.
\end{highlights}


\end{frontmatter}



\section{Introduction}
\label{introduction}

The CLAS12 detector system at Jefferson Lab, Virginia, measures high energy collisions of electrons and nucleons to advance our understanding of fundamental nuclear physics. Similar to many other particle physics experiments, CLAS12 uses electromagnetic calorimeters (ECals) \cite{Burkert_Elouadrhiri_Adhikari_etal._2020} to help measure the type, position and energy of these final state particles.

\begin{figure}[htbp]
    \centering
    \includegraphics[width=0.9\textwidth]{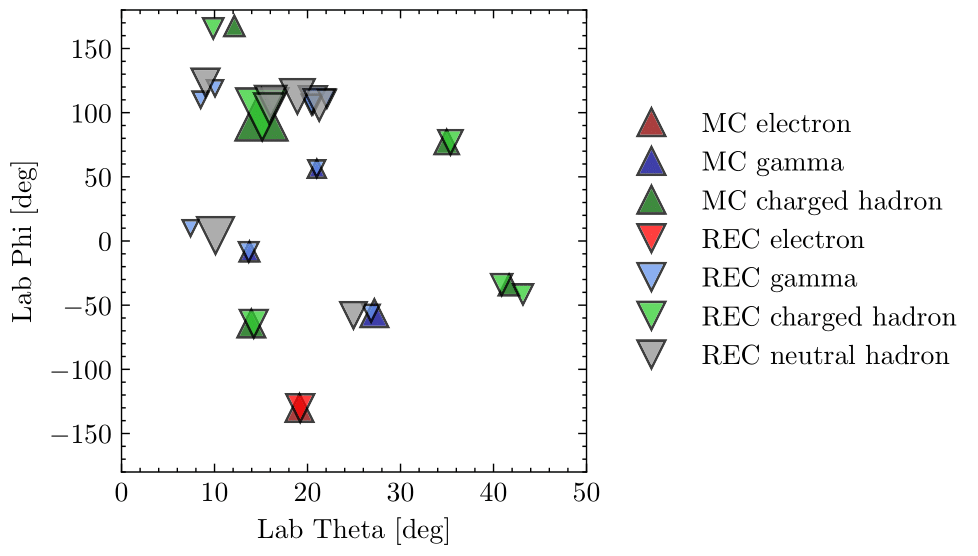}
    \caption{Sample $e^-+p$ Monte Carlo event at CLAS12 reconstructed using \coatjava plotted in $(\theta,\phi)$ space. Here, $\theta$ and $\phi$ are defined in the lab-frame where the proton target is at rest. Upwards (downwards) facing triangles represent the final state true (reconstructed) particles. Marker size roughly scales with particle energy.}
    \label{fig:particle_map}
\end{figure}

The CLAS12 ECal is important for measuring photons and neutrons. These particles do not leave tracks in the tracking detectors and can therefore only be identified using their energy deposits and timing information in the calorimeters. At CLAS12, in the absence of a dedicated hadronic calorimeter, neutrons and photons are primarily distinguished by comparing the event's trigger time with the timing of calorimeter hits as neutrons arrive later than photons due to their slower speeds. Physics studies using photons from $\pi^0$ decays or neutrons from exclusive events thus rely on the ECal for accurate reconstruction of these particles. Inefficiencies in the collaboration's analysis pipeline \coatjava lead to an overabundance of \textit{fake} neutral particles being reconstructed, complicating these studies. These fake particles introduce background that contaminate genuine signals, making it challenging to isolate and study the true neutral particle production mechanisms. To illustrate this, Figure \ref{fig:particle_map} shows the $(\theta,\phi)$ of Monte Carlo generated particles for a sample $e^-+p$ event and compares with those reconstructed using \coatjava. Whereas a perfect detector system would reconstruct a single particle for each generated, issues in \coatjava's calorimeter reconstruction algorithm lead to a clear overabundance of false neutral particles.

As a result, CLAS12 analyses that rely on neutrons may adopt conservative selection criteria to suppress false tags at the cost of reducing resolution and statistics. Exclusive channels such as $\pi^+/\rho^+$ production off the proton ($e^-+p\rightarrow e^-+\rho^+/\pi^++n$), $J/\psi$ production off deuterium ($e^-+d\,(n)\rightarrow e^-+J/\psi+n'$), and Deeply Virtual Compton Scattering on the neutron ($e^-+n\rightarrow e^-+\gamma+n'$) utilize missing mass/momentum cuts to identify the event's neutron with appropriate kinematics, yet are still prone to selecting false positives \cite{Tyson2023,Diehl_2023,PhysRevLett.133.211903}. A handful of neutron-related physics channels that fall primarily within the ECal's acceptance are simply impossible due to the existing reconstruction efficiency. These processes, for instance semi-inclusive neutron production ($e^-+p\rightarrow e^-+n+X$) and back-to-back dihadron production (ex: $e^-+p\rightarrow e^-+\pi+n+X$), would allow CLAS12 to complete the full spectrum of SIDIS measurements with all flavor combinations in the target- and current-fragmentation regions \cite{PhysRevLett.130.022501,kripko2025multidimensionalmeasurementsbeamsingle,Diehl_2022}.

In this work, we propose an AI model that significantly enhances the clustering accuracy of ECal hits at CLAS12. The network learns hit positional encodings using a module composed of consecutive GravNet layers \cite{Qasim_Kieseler_Iiyama_etal._2019} which process the complete ECal detector topology. The learned encodings are combined with the embedded hit representations for each event, and the resulting feature vector is then processed by a Transformer encoder \cite{Vaswani_Shazeer_Parmar_etal._2023}. Then, a multi-layer perceptron (MLP) network clusters tokens (hits) belonging to the same particle by mapping them to similar regions in an 2-dimensional latent space. This clustering strategy, known as object condensation (OC), was first introduced in Ref. \cite{Kieseler_2020} and has since been employed in several nuclear and particle physics AI clustering methods \cite{Lieret_DeZoort_Chatterjee_etal._2023, Calafiura}. To our knowledge, this work represents the first instance of AI-assisted clustering applied to hodoscopic detectors. Unlike pixelated, granular detectors, hodoscopes determine particle positions from intersecting "trip-wire" planes arranged in a criss-cross geometry. Another novelty is that the network combines fine-grained local neighborhood representations (GravNet) with long range contextual information (self-attention) to strengthen event reconstruction.

The paper is organized as follows: Section \ref{sec:rel} gives an overview of machine learning applications in particle and nuclear physics, with a focus on clustering tasks. Section \ref{sec:clas} describes the CLAS12 electromagnetic calorimeter and the current \coatjava clustering method. In Section \ref{sec:data}, we describe the event simulation and dataset. Section \ref{sec:model} showcases the new model architecture for performing ECal hit clustering, and discusses the object condensation loss. In Section \ref{sec:results}, the results of the model are shown, and a new metric is defined to compare with \coatjava. We summarize our findings and detail future work in Section \ref{sec:summary}.

\section{Related Work}\label{sec:rel}
Machine learning applications in particle and nuclear physics are constantly evolving, with tasks ranging from clustering, identification, regression, and fast simulation. A living review containing hundreds of these applications can be found in Ref. \cite{hepmlLivingReview}. Graph neural network (GNN) based architectures form the backbone of the majority of detector-based clustering tasks. This is due to their ability to represent irregular detector topologies with a flexible learned latent representation \cite{10.1145/3326362, Wemmer_Haide_Eppelt_etal._2023}. We separate this discussion to remark on the current progress of AI approaches to track and calorimeter clustering.

\subsection{Track clustering}

The Exa.TrkX collaboration's \textbf{GNN4ITk} \cite{Caillou:2815578} pipeline performs track clustering by first constructing a graph where nodes represent hits in the silicon inner tracker. The GNN edges are scored to assign low/high probability connections. After edge filtering, track candidates are collected in an iterative graph segmentation stage. GNN4ITk's edge classification framework became a leading approach, inspiring further improvements and models for addressing the high multiplicity challenges of the future HL-LHC.

Hierarchical Graph Neural Networks \textbf{HGNNs} \cite{Liu_Calafiura_Farrell_etal._2023} addresses the difficulty of handling disconnected tracks with GNN track clustering. Initial connectivity during graph construction may prohibit a disconnected track from being properly reconstructed. In HGNN, track segments are pooled into \textit{super-nodes}, at which a k-NN operating on super-nodes allows message passing between broken segments, increasing the receptive field and preserving long-range relationships. This in turn allows the model to learn to combine disconnected track segments to form one track. 

The LHCb collaboration's \textbf{ETX4VELO} \cite{Correia_Giasemis_Garroum_etal._2024} model builds on GNN4ITk, addressing another major concern with GNN track clustering --- node sharing tracks. Identifying two tracks tending to the same shared node(s) is solved by introducing an additional classifier that operates on a graph of the edges. In a \textit{triplet-building} stage, the model learns to duplicate nodes belonging to separate tracks, splitting the network into multiple tracks, making clustering more straightforward. 

The Evolving Graph-based Graph Attention Network \textbf{EggNet} \cite{Calafiura} avoids explicit initial graph construction by allowing the nodes to learn edges connections dynamically. When the final edge connections are proposed, each contributes some amount to an object condensation based loss term based on if the two connected nodes belong to the same particle. This approach enhances message passing, improves graph efficiency, and mitigates issues related to missing track connections.

Transformer-based models use more modern architectures compared with traditional GNN methods. One such model \cite{Stroud_Duckett_Hart_etal._2024} uses the \textbf{MaskFormer} architecture --- originally developed for image segmentation \cite{Cheng_Schwing_Kirillov_2021} --- to simultaneously assign hits to tracks and predict track properties. The approach begins with a Transformer encoder with a sliding window to filter hits by classifying them as signal or noise. The signal hits are then processed through a fixed-window encoder-decoder module that generates multiple binary masks corresponding to hit-to-track assignments. The model was tested on the TrackML dataset with great success and shows a growing trend for integrating computer vision-inspired solutions to clustering.








\subsection{Calorimeter clustering}

A \textbf{fuzzy-clustering GNN} \cite{Wemmer_Haide_Eppelt_etal._2023} was developed for the Belle II experiment to address overlapping photon showers from $\pi^0$ decays in their electromagnetic calorimeter. In this work, each calorimeter crystal performs message-passing in a dynamically generated graph using GravNet \cite{Qasim_Kieseler_Iiyama_etal._2019}. The GNN predicts a set of weights that determine the fractional assignment of each hit to multiple potential clusters, allowing for partial energy contributions to overlapping photon showers. The study outperforms the baseline Belle II reconstruction algorithm, achieving a 30\% improvement in energy resolution for the low energy photons in asymmetric photon pairs. 

An \textbf{object condensation-based GravNet} approach \cite{Qasim_Long_Kieseler_etal._2021} was developed for the CMS High Granularity Calorimeter (HGCAL) at the future HL-LHC, where up to 200 simultaneous proton-proton interactions may occur. Similar to Belle II, due to the irregular geometry of the HGCAL, GravNet's flexibility in assigning nearest neighbors allows for efficient clustering. For each hit, the network predicts object condensation variables for clustering, as well as cluster properties such as particle energy. This study demonstrates the potential for end-to-end multi-particle reconstruction at the HL-LHC.

\section{The CLAS12 Electromagnetic Calorimeter}\label{sec:clas} 

\begin{figure}[ht]
    \centering
\includegraphics[width=0.95\textwidth]{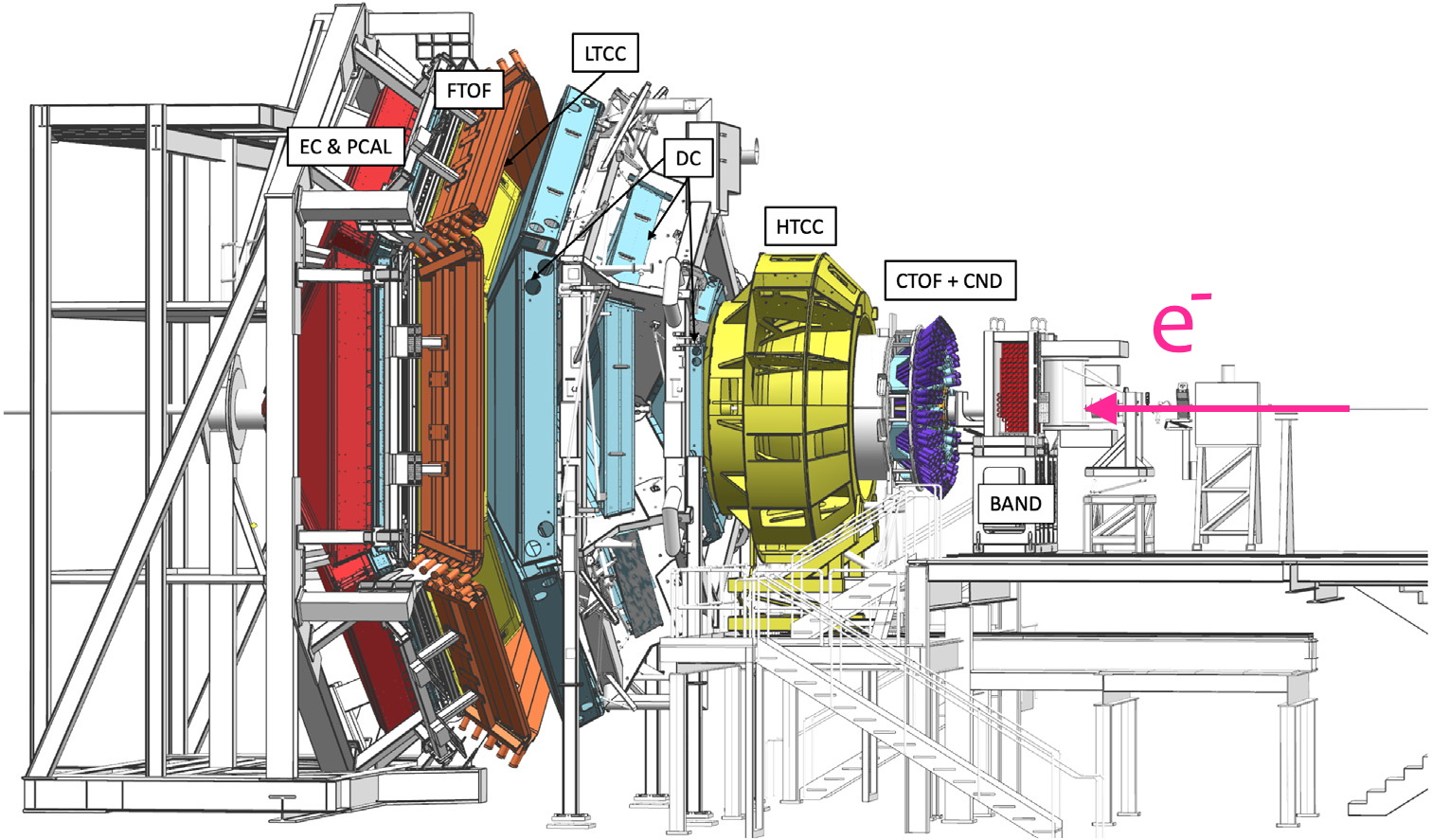}
    \caption{A schematic of the CLAS12 detector system \cite{Burkert_Elouadrhiri_Adhikari_etal._2020}. The proton target is positioned within the Central Neutron Detector (CND) and central time-of-flight (CTOF), which are downstream from the backward angle neutron detector (BAND). The electromagnetic calorimeter consists of the PCal, ECin and ECout (the latter two grouped as EC in the figure) subsystems. The forward-angle detector system exhibits six-fold azimuthal symmetry, evident by the distinct corners of its EC, PCal, forward time-of-flight (FTOF), and drift chamber (DC) detectors. The high threshold Cherenkov counter (HTCC) and low threshold Cherenkov counter (LTCC) help discriminate charged particles.}
    \label{fig:clas12}
\end{figure}

\begin{figure}[htbp]
    \centering
    \begin{subfigure}[b]{0.45\textwidth}
        \centering
        \includegraphics[width=\textwidth]{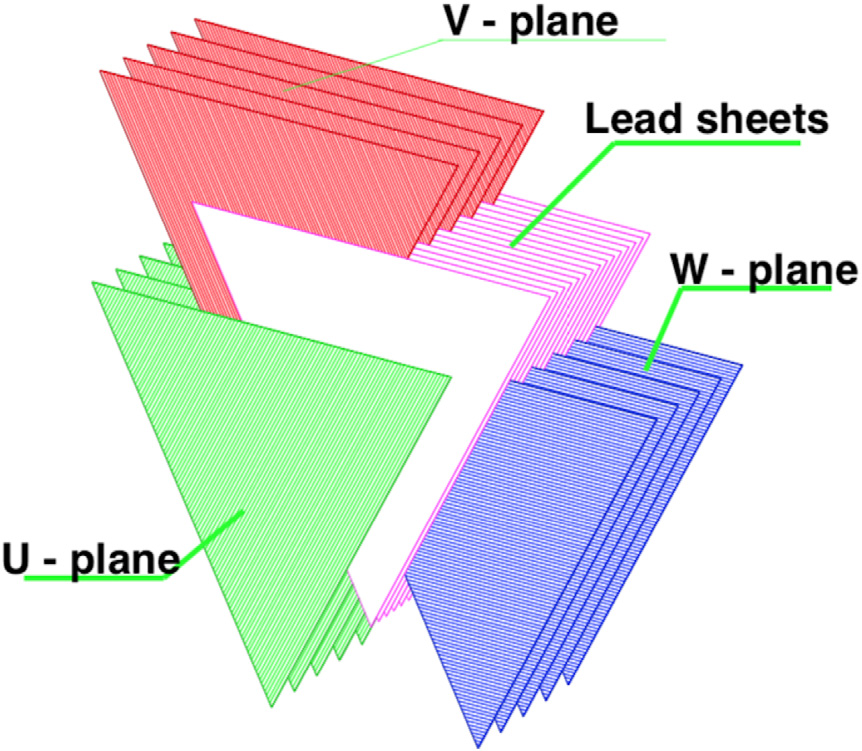}
        \caption{Three dimensional geometry of a single PCal sector (exploded view).}
        \label{fig:ecalpic1}
    \end{subfigure}
    \hfill
    \begin{subfigure}[b]{0.45\textwidth}
        \centering
        \includegraphics[width=\textwidth]{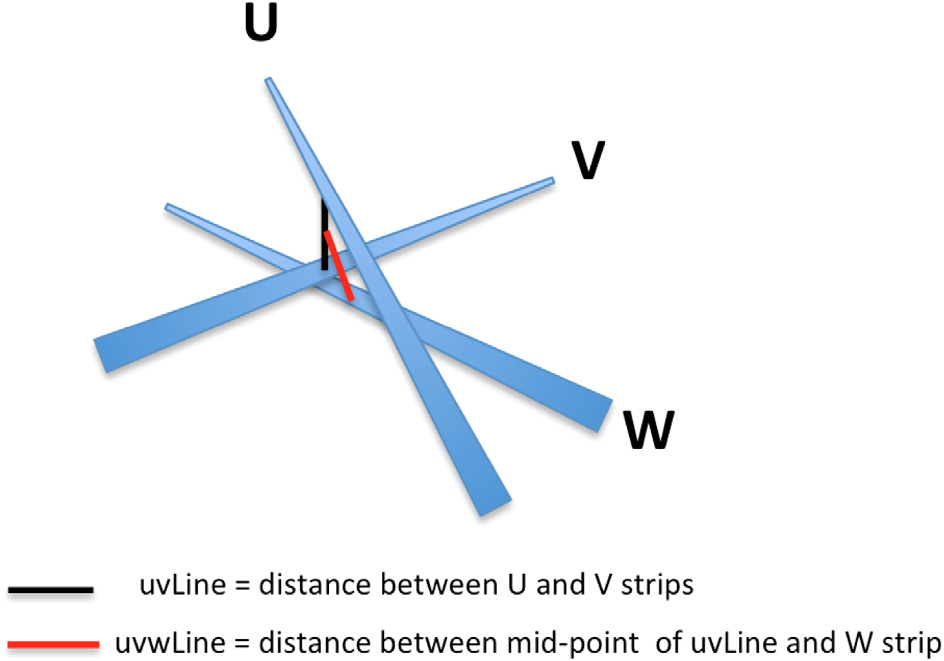}
        \caption{Schematic of a 3-way intersection of U, V, and W strips. The midpoint of the \textit{uvwLine} defines the cluster coordinates.}
        \label{fig:ecalpic2}
    \end{subfigure}
    \caption{View of the U, V, W scintillating strip plane layout at CLAS12 \cite{Asryan_Chandavar_Chetry_etal._2020}.}
    \label{fig:ecalpics}
\end{figure}

CLAS12's forward detector system consists of 6 distinct azimuthal sectors arranged around the beam pipe (see Figure \ref{fig:clas12}). The laboratory frame uses a right-handed coordinate system, with the z-axis aligned along the direction of the electron beam and the x- and y-axes defined accordingly. Each sector contains three calorimeter subsystems, named the pre-shower calorimeter (PCal), inner calorimeter (ECin)
and the outer calorimeter (ECout), ordered in increasing distance from the collision point. Each subsystem is a sampling calorimeter comprised of $1$cm thick scintillator strips and $1/4$cm thick lead sheets, separated by a $50\mu\mathrm{m}$ Teflon sheet \cite{Asryan_Chandavar_Chetry_etal._2020} (see Figure \ref{fig:ecalpic1}). The three calorimeter subsystems combine to form approximately 22 radiation lengths of material. Each of the six sectors' PCal contains 192 scintillator strips, the ECin 108 strips, and the ECout 108 strips, for a total of 2448 for the whole system. Scintillator strips are arranged in a triangular layout such that each layer of strips (named U, V, and W) are $\pm120^\circ$ relative to one another.

As for its design choice, the ECal was engineered to satisfy the following physics requirements:
\begin{itemize}
  \item Electron and photon energy resolution of  
    \[
      \frac{\sigma_E}{E} \;\le\; \frac{0.1}{\sqrt{E\;[\mathrm{GeV}]}}
    \]
  \item Shower position resolution of \(\approx 1\ \mathrm{cm}\)
  \item Pion misidentification rate below \(1\%\) for \(E_{\pi}\ge5\ \mathrm{GeV}\)
  \item Invariant mass resolution for \(\pi^0\!\to2\gamma\) decays satisfying  
    \(\delta m/m \le 0.1\)
  \item Neutron detection efficiency exceeding \(50\%\) for \(E_n>1\ \mathrm{GeV}\)
  \item Time‐of‐flight precision on the order of \(0.5\ \mathrm{ns}\)
\end{itemize}
CLAS12's addition of a high-resolution PCal from CLAS6's original design was necessary to capture small opening angle $\pi^0\rightarrow\gamma\gamma$ decays following a doubling of the electron's beam energy (from $6$ GeV up to a maximum of $12$ GeV). 

Subatomic particles from the collision generate secondary particles, primarily from interactions in the lead sheets. These secondaries then pass through the scintillator strips which emit light in response. The light propagates throughout the strip and is read out by photomultiplier tubes (PMTs) to record deposited energy. The individual reading of a strip in a collision event is referred as a \textit{hit}. A \textit{cluster} is a group of hits from the same initial particle. 

Clusters are essential objects defined during event reconstruction that capture the position and total energy deposited by a particle. Collider experiments such as CMS \cite{Collaboration_2017}, ATLAS \cite{ATLAS:2010cba}, and ALICE \cite{Collaboration_2023} have grid-like calorimeter topologies and use a seeding algorithm to group hits into clusters. To form clusters at CLAS12, adjacent strips within the same layer (ex: U) are first collected into intermediate objects called \textit{peaks}. In a process sketched in Figure \ref{fig:ecalpic2}, \coatjava determines a cluster using geometry by searching for 3-way intersections of U, V, W peaks. Besides CLAS12, ECals with hodoscopic geometries (ones that exploit cross-layered strips for determining cluster position using intersections) are seen in a range of physics experiments \cite{Allan_Andreopoulos_Angelsen_Barker_Barr_Bentham_Bertram_Boyd_Briggs_Calland_etal._2013, Collaboration_Atwood_2009, 10.1063/1.4919462}.

\begin{figure}[ht]
    \centering
\includegraphics[width=0.95\textwidth]{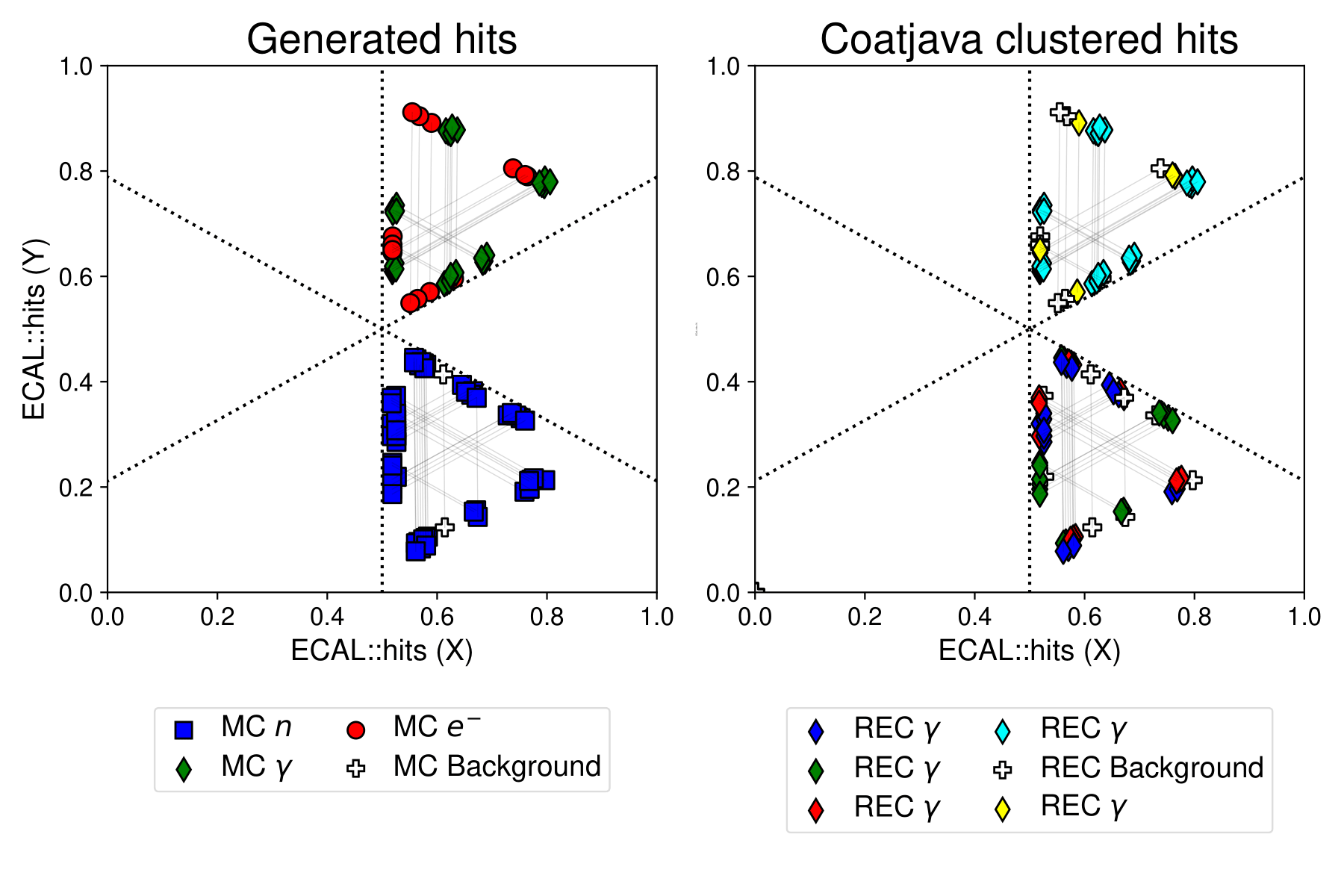}
    \caption{Simulated ECal detector response from multiple final state particles in a sample collision event. For each plot, colors represent distinct particles and marker-styles label particle types ($e^-$, $\gamma$, etc). (Left) Scintillator strips are labeled with the truth-level SIDIS particle which caused the hit. (Right) Strips are labeled by the distinct clusters (particles) reconstructed by \coatjava. In the bottom right sector, a true generated neutron deposits energy in many strips. When processing this group of hits, the \coatjava clustering algorithm reconstructs three neutral particles and misidentifies them as photons. The electron generated in the top-right sector is misidentified as a photon due to a missing drift chamber track match.}
    \label{fig:ecal_hits_simple}
\end{figure}

An important question to address is how does \coatjava reconstruct \textit{fake} neutral particles in the first place? For each cluster of ECal hits measured at CLAS12, \coatjava looks for a track whose trajectory points towards the cluster centroid. Clusters without a matching track are classified as neutrals, and timing information, among other properties of the cluster, are used to distinguish between neutrons and photons. Fake neutral particles are most often reconstructed because \coatjava incorrectly interprets the hits generated by one true particle as multiple independent clusters. We see this happen in sample events such as one shown in Figure \ref{fig:ecal_hits_simple}. In this event, the widely dispersed ECal hits from a single Monte Carlo neutron causes three separate neutral particles to be reconstructed. These reconstruction issues are able to be overcome through the design of our AI clustering model.

\section{Dataset}
\label{sec:data}

In this work, one million $e^-+p$ fixed target Deep Inelastic Scattering (DIS) events (collisions) are generated using \textsc{clasdis}, a Semi-inclusive DIS (SIDIS) Monte Carlo based on PEPSI LUND \cite{clas12_mcgen}. The electron beam energy was set to $10.6$ GeV to replicate the configuration of previous, ongoing, and future CLAS12 experiments. Final state particles for each event are processed using a Geant4 Monte Carlo simulation framework called \textsc{gemc} to create realistic detector readouts from CLAS12. For each event, the 150 ECal strips with the highest energy deposits are selected, and each is assigned 17 features-with zero-padding applied if fewer than 150 strips are hit. For each strip, the features are:
\begin{itemize}
    \item The Cartesian coordinate endpoints of the strip, labeled $x_o,y_o,z_o$ and $x_e,y_e,z_e$.
    \item The energy deposited in the strip.
    \item The timing recorded by the strip.
    \item 9 one-hot encoded bits to assign the strip's layer number. There are 3 calorimeters (PCal, ECin, ECout) and a U,V,W layer for each.
\end{itemize}

All features, such as the timing information, are scaled between $[0,1]$ to avoid exploding gradients during training.

An additional feature per strip, its \textsc{stripID}, uniquely identifies it among the 2448 strips in the CLAS12 ECal. The \textsc{stripID} is utilized in the model to cross-reference the strip coordinates and to match to correct positional encodings.

To properly train the clustering algorithm, hits belonging to the same particle in the event are assigned a unique true ID. To do so, the particle history of each ECal hit in an event is traced back in Geant4 to one of the final state particles generated in the collision. All zero-padded hits are assigned a true ID of -1 to mark them as background. In a pre-processing step, we check each sector's PCal, ECin, and ECout for at least one hit all three layers --- U,V and W. If two or less are found, then those hits are considered background as later postprocessing steps require all three to form a cluster.

\section{Methodology}
\label{sec:model}
\subsection{Architecture}
\begin{figure}[htbp]
    \centering
    \begin{subfigure}[b]{\textwidth}
        \centering
        \includegraphics[width=\textwidth]{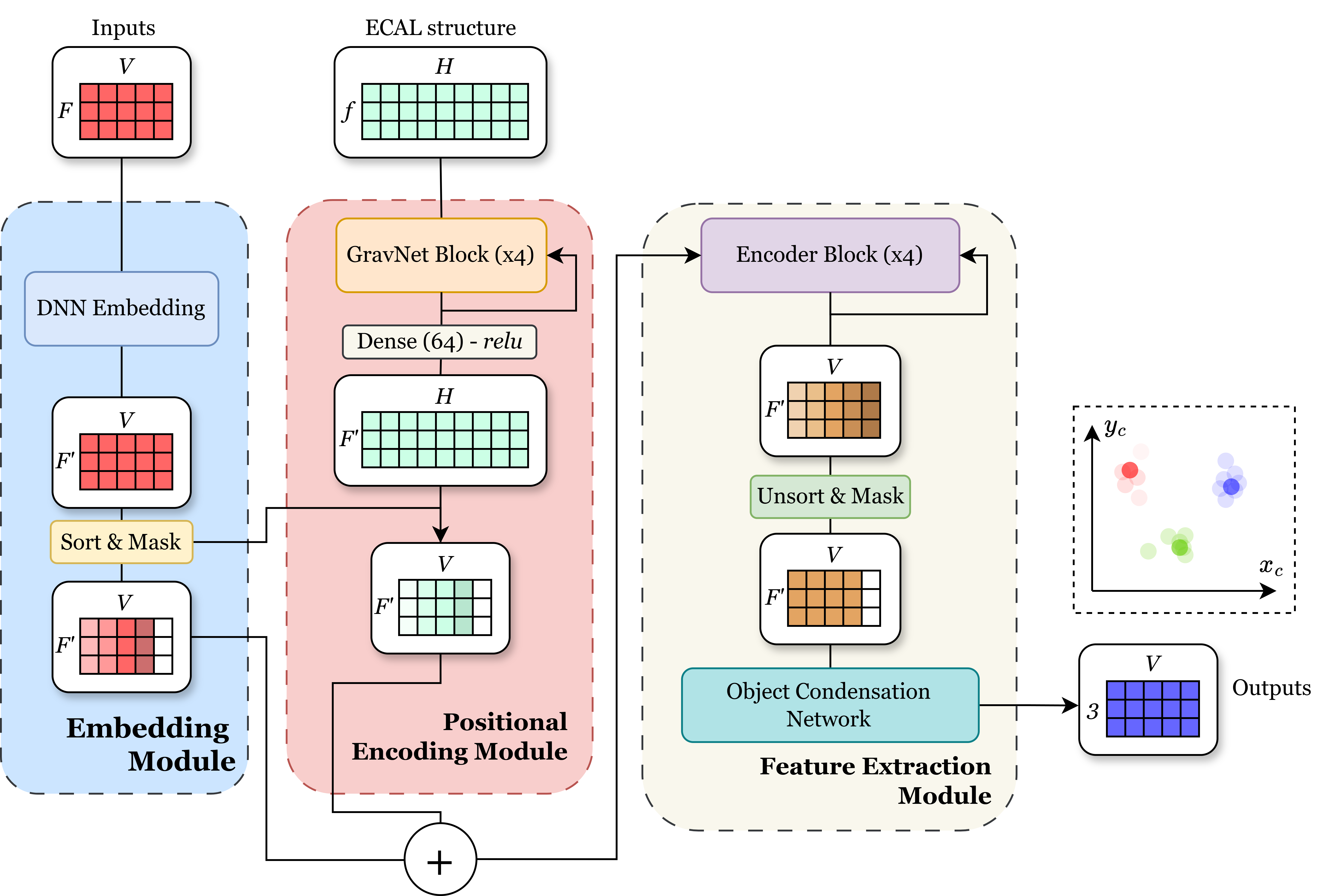}
        \caption{High-level view of the full architecture pipeline. Event-by-event ECal hits are inputted into the embedding module, preparing them for the encoder blocks. A positional encoding module trains on the entire detector topology, adding these encodings to the hits. A feature extraction module predicts object condensation variables for each hit for later clustering.} 
        \label{fig:model.va}
    \end{subfigure}
    
    \vspace{1em} 
    
    \begin{subfigure}[b]{\textwidth}
        \centering
        \includegraphics[width=\textwidth]{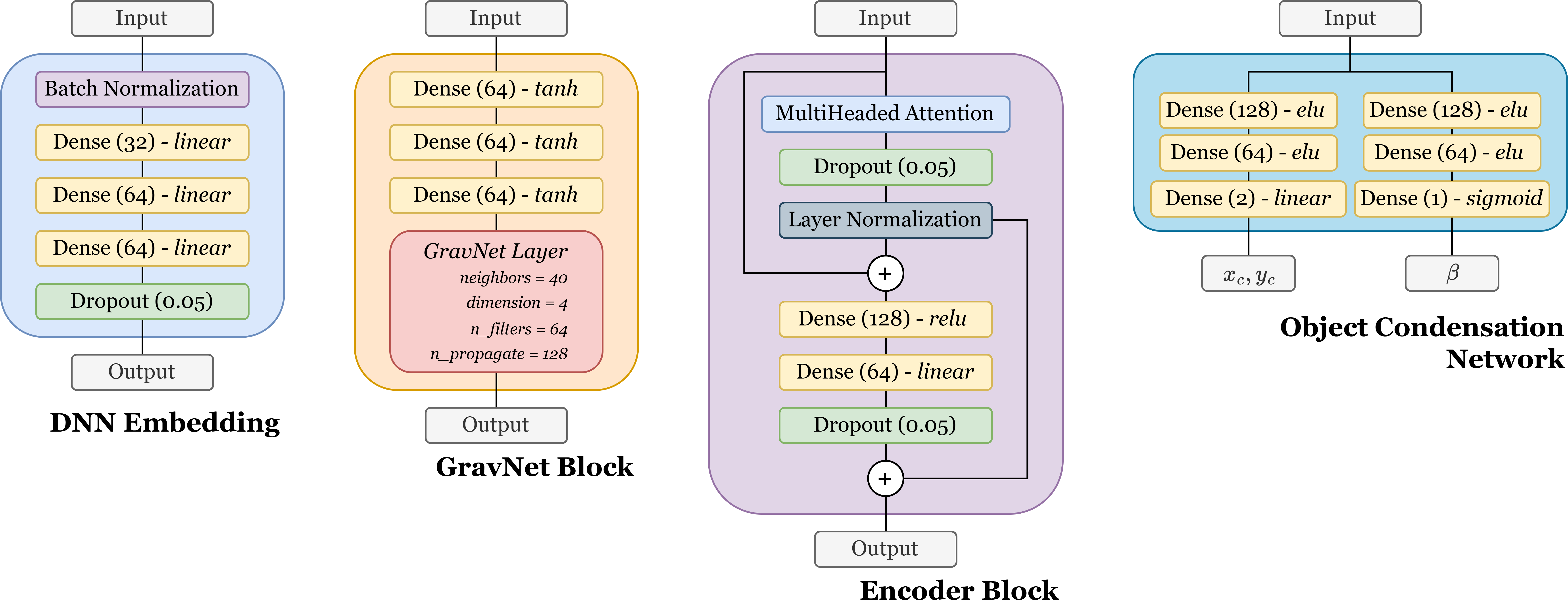}
        \caption{Internal components of the networks modules. For this project, we utilize the Python library TensorFlow to construct each layer.}
        \label{fig:model.vb}
    \end{subfigure}
    
    \caption{Schematic of the network architecture, highlighting: (a) the higher-level embedding, positional encoding, and feature extraction modules, (b) the sub-networks built into these modules. }
    \label{fig:model}
\end{figure}

The model architecture is illustrated in Figure \ref{fig:model}. The network is comprised of three modules --- embedding, positional encoding, and feature extraction. The input is a point cloud $x\in\realnumbers^{V\times F}$ consisting of $V=150$ ECal hits, where each point is represented by $F=17$ features.

The point cloud $x$ is passed through the embedding module $f_{\mathrm{EMB}}$. Input node features are passed through a batch normalization layer and are then encoded using 3 MLPs with linear activations, followed by a 0.05 feature dropout (see Figure \ref{fig:model.vb}). We then sort the output along the $V$-dimension by the \textsc{stripID}, saving the unmasked strip hits for the positional encoding. The output $z$ of the embedding module is given by

\begin{equation}
z\in\realnumbers^{V\times F'}=f_{\mathrm{EMB}}(x),
\end{equation}
where $F'=64$. 

At the same time, a positional encoding (PE) module $f_{\mathrm{PE}}$ receives as input $g\in\realnumbers^{H\times f}$ that represents the full detector topology. The $f=6$ coordinate features representing each of the $H=2448$ strips are the $(x,y,z)$ of the two strip endpoints. This fixed tensor is passed through 4 consecutive GravNet blocks, each containing 3 MLPs followed by a single GravNet layer. Although the embedding module already projects raw hit features into a higher‐dimensional space, a pure Transformer treats its inputs as an unordered set. In our case the order of the 2448 strips encodes real, fixed topology information (e.g. strip \#105 is adjacent to strip \#106, or strip \#4 intersects with strip \#95). Without positional signals, the network could not tell whether two hits came from neighboring strips or from opposite sides of the detector. By learning a 64-dimensional positional embedding for each strip, we guarantee that each hit carries both its local features \textit{and} its absolute (and relative) location in the calorimeter.

Following the works of \cite{Wemmer_Haide_Eppelt_etal._2023, Qasim_Long_Kieseler_etal._2021}, we chose GravNet \cite{Qasim_Kieseler_Iiyama_etal._2019} due to its ability to perform message-passing in learned latent graphs without explicit construction.  In each GravNet layer, every node is mapped to a latent $S$-space where it learns intrinsic features. Each node’s 
$S$-space representation is then refined by aggregating information from its $k$-nearest neighbors using distance-weighted mean and max functions, and these aggregated features, along with the node’s original and $S$-space features, are updated via a fully-connected MLP. 

The resulting features from each GravNet block are concatenated and passed through a single MLP to obtain $F'$ hidden features for each strip. The new hidden geometry representation is truncated, masked, and sorted to align with the non-background \textsc{stripID}s of the embedding module's output $z$. The positional encoding module's output $g'$ is defined as 

\begin{equation}
    g'\in\realnumbers^{V\times F'}=f_{\mathrm{PE}}(g;x),
\end{equation}
and is added to the embedding module's output $z$ to serve as the input to the feature extraction module.

The feature extraction module $f_{\mathrm{EXT}}$ is composed of a Transformer encoder and a dense network for determining object condensation variables. The $V=150$ length sequence of $F'$-dimensional tokens are passed through a background-masked self-attention mechanism. Because we always supply a fixed‐length sequence of 150 tokens into the Transformer encoder, but only a small subset correspond to actual hits, we apply a binary mask that zeroes out all empty positions in both the attention scores and the positional embeddings. This masking allows every hit to attend to all others, capturing long-range relationships such as multi-sector clusters. The self-attention layer is followed by dropout and layer normalization. The resulting representation is then concatenated with the original sequence and forwarded through two fully-connected MLPs, after which another dropout is applied. A skip connection, combining the layer normalization output to the final dropout, gives the output of a single encoder block. After 4 consecutive encoder blocks, the sequence's sorting is reversed to reflect the original ordering of the model's inputs. The features of background hits are zeroed once more. 

The hit representation is passed through a final dense network that determines a 2-dimensional latent space coordinate ($x_c,y_c$) and confidence measure $\beta$ for each hit, such that:

\begin{equation}
    y\in\realnumbers^{V\times 3}=f_{\mathrm{EXT}}(z+g').
\end{equation}

\subsection{Loss Function}

The full network maps each ECal hit to a location in a 2-dimensional latent space and assigns it a confidence value between $\beta\in[0,1]$. High values of $\beta$ indicate a stronger condensation point. In the latent space, condensation points attract other hits belonging to the same cluster, and repel hits that belong to other clusters. To reinforce this behavior, Ref. \cite{Kieseler_2020} describes an attractive and repulsive potential loss $\mathcal{L}_V$. First, for each true cluster $t$, the hit with the highest $\beta$ is named that cluster's representative. The attractive loss is quadratic in the distance between a hit and its cluster's representative, pulling it towards the condensation point. The repulsive loss is linear and repels hits from condensation points of other objects. 

An additional loss term, the beta loss $\mathcal{L}_\beta$, helps tune the value of $\beta$ for the hits. It is comprised of two components, the first being the "coward loss" which rewards the network for maximizing the $\beta$ of condensation points. The second component, the "noise loss", penalizes the model for assigning high $\beta$ to noisy hits. The total object condensation loss to minimize is thus
\begin{align}
    \mathcal{L}&=\mathcal{L}_V+\mathcal{L}_\beta\nonumber\\
               &=\mathcal{L}_{\mathrm{att}}+\mathcal{L}_{\mathrm{rep}}+\mathcal{L}_{\mathrm{cow}}+\mathcal{L}_{\mathrm{nse}},
\end{align}
where $\mathcal{L}_{\mathrm{att}}$ is the attractive loss, $\mathcal{L}_{\mathrm{rep}}$ is the repulsive loss, $\mathcal{L}_{\mathrm{cow}}$ is the coward loss and $\mathcal{L}_{\mathrm{nse}}$ is the noise loss.

\begin{figure}[htbp]
    \centering
    \includegraphics[width=\textwidth]{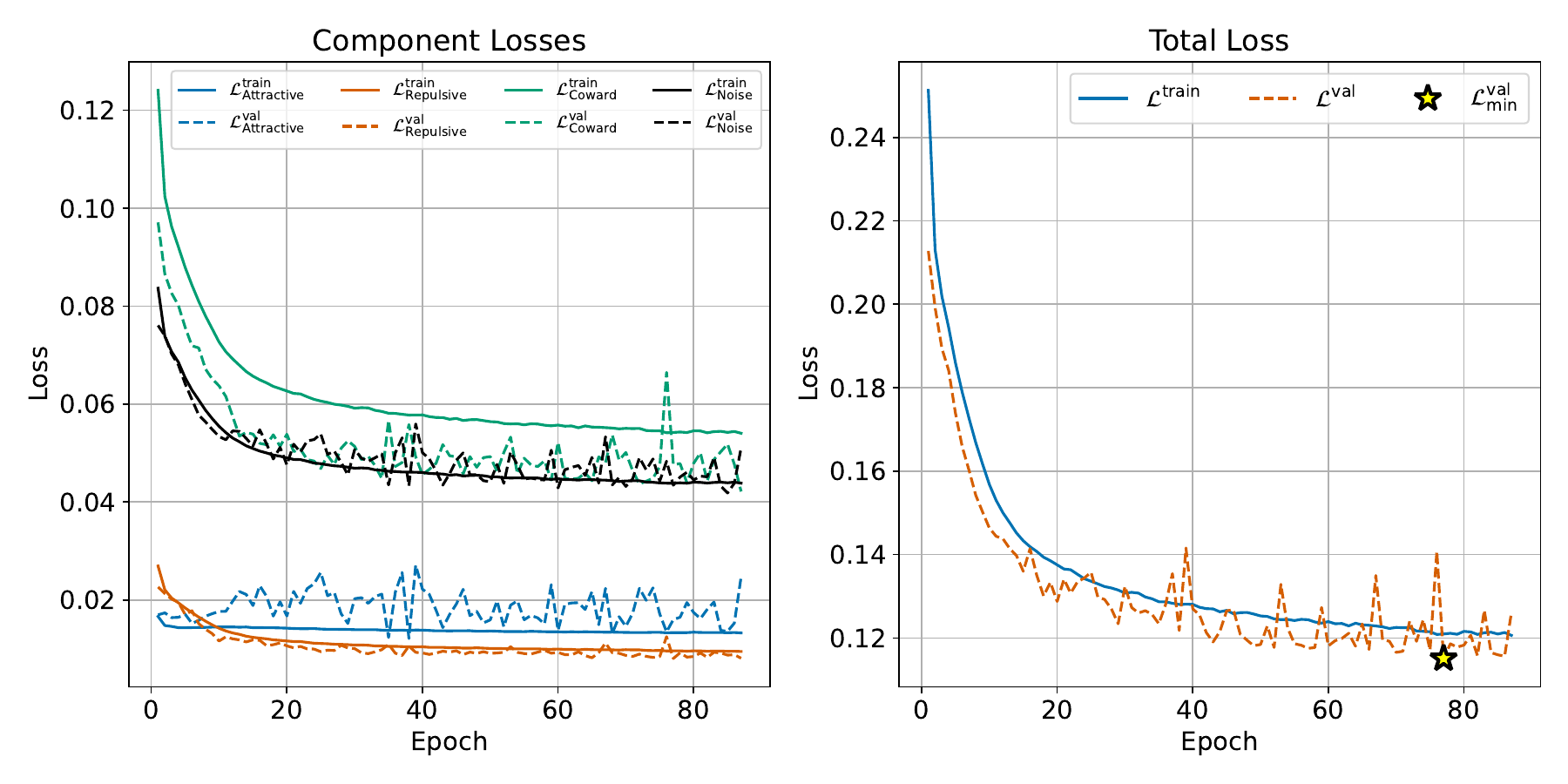}
    \caption{(Left) Components of the object condensation loss as a function of training epoch. (Right) Sum of the object condensation loss components as a function of training epoch. Dashed lines represent the validation loss.}
    \label{fig:loss}
\end{figure}

Figure \ref{fig:loss} shows how the training and validation losses evolve over the epochs. Note that dropout layers are disabled during validation, which explains the lower loss calculated on the validation set compared to the training set. We selected the final model parameters from epoch 77, as no further improvement was seen over the subsequent 10 epochs.

\subsection{Inference}

\begin{figure}[htbp]
    \centering
    \includegraphics[width=0.9\textwidth]{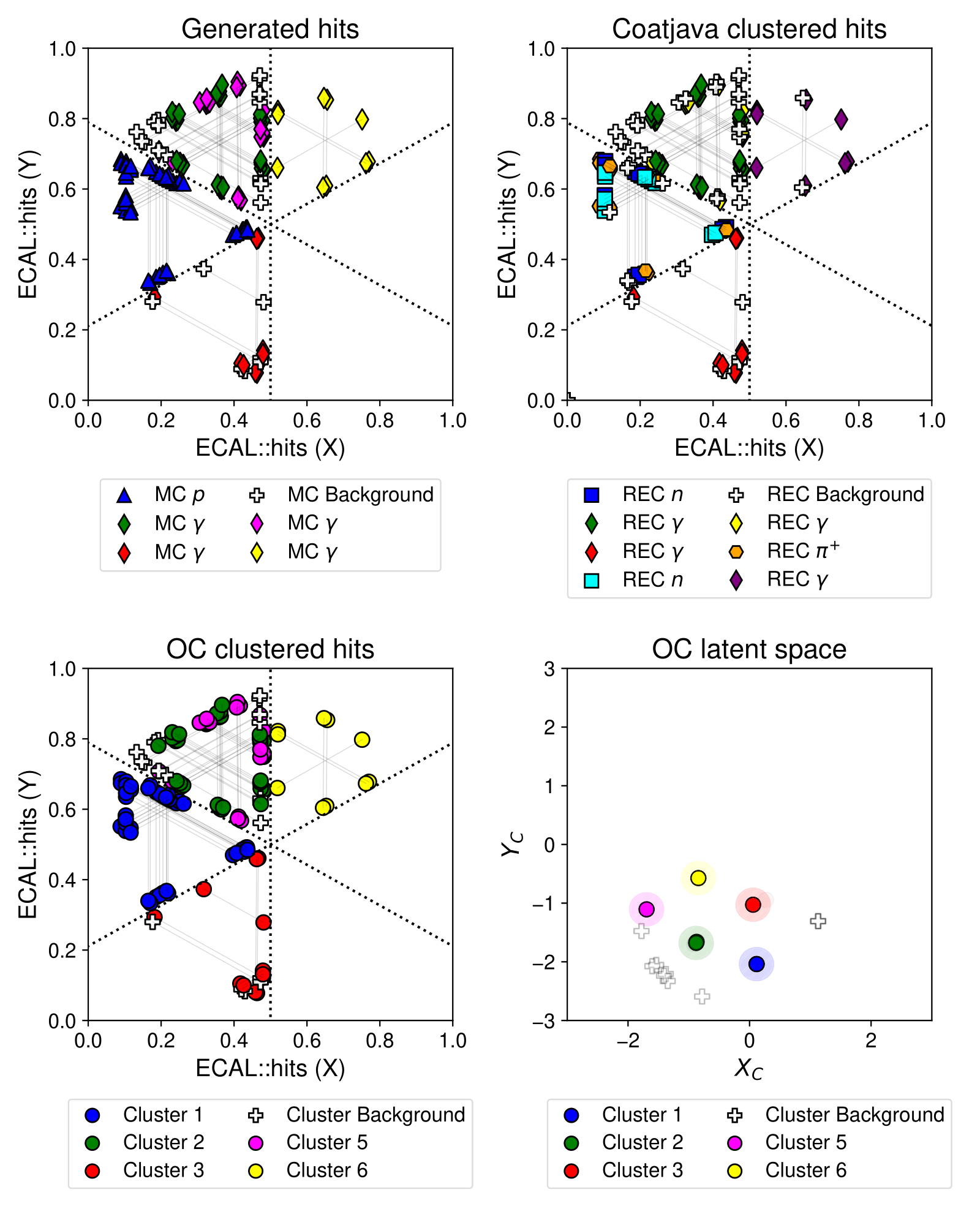}
    \caption{(Top Left) ECal hits left by generated SIDIS particles in a collision event. (Top Right) The clusters and particles reconstructed by \coatjava after interpreting the generated hits. (Bottom Left) The clusters reconstructed by our Object Condensation model from the same generated ECal hits. (Bottom Right) A snapshot of the clustered 2-D latent space from which our Object Condensation model maps the generated hits.}
    \label{fig:oc_example}
\end{figure}

Clustering starts by first ordering all hits by their learned $\beta$. To create the first cluster, the highest $\beta$ is chosen, and all hits within a distance $t_D=0.28$ of it are grouped together. Then, the second cluster begins with the next highest $\beta$ that is unclustered. We repeat this iteratively, forming clusters until the highest remaining $\beta$ falls below the threshold $t_\beta=0.5$. All remaining unclustered points are classified as noise. 

In Figure \ref{fig:oc_example}, we compare the ECal clustering using \coatjava and object condensation for a sample event's detector response. Generated hits are mapped to locations in the OC latent space where the previously mentioned inference steps are taken to define OC clusters. In this example, a single particle, a proton, entered into the leftmost sector, creating a handful of ECal hits. After reconstruction, \coatjava infers three separate particles created these hits, two of which are false neutrons, and the other a mis-identified $\pi^+$. This result indicates \coatjava's clustering algorithm located three clusters where there should only be one. In comparison for the same event, the object condensation model correctly identifies only one cluster in the leftmost sector. Furthermore, the model correctly identifies the two distinct clusters in the top-left sector, whereas \coatjava only finds one. In the latent space, these clusters are well separated, indicating the model's confidence that these group of hits indeed belong to separate objects. 

\subsection{Postprocessing}
\label{postprocessing}

A postprocessing step transforms each group of strips into an ECal cluster object. These objects are sent through the rest of the \coatjava reconstruction pipeline, allowing us to bypass the previous clustering algorithm.

Determining the cluster centroid is a two-step process. The first step collects $N$ 3-way intersections for each cluster $k$. The second step uses those $N$ 3-way intersections to calculate one cluster centroid for each cluster $k$. In more detail\ldots

\begin{enumerate}
    \item \textbf{Loop over PCal, ECin, and ECout strips}
    \begin{itemize}
        \item For each strip \( j \) belonging to a cluster \( k \), find its most energetic \( \left( \sum_j E_j \right) \) 3-way intersection.
        \item A 3-way intersection is defined by the average \((x, y, z)\) of the closest approach for strips \( uv \), \( vw \), and \( uw \).
        \item The energy \( E_j \) for each strip is corrected to account for attenuation.
    \end{itemize}
    
    \item \textbf{For each cluster \( k \) containing \( N \) 3-way intersections}
    \begin{itemize}
        \item Only consider 3-way intersections in the sector with a 50\%+ majority.
        \item Calculate the z-score \( z_i \) for each 3-way intersection \((x, y, z)\).
        \item Report the centroid’s \((x, y, z)\) as the weighted sum of the 3-way intersections, where \( w_i = (1 + z_i^2)^{-1} \) to lessen the impact of distantly separated 3-way intersections.
    \end{itemize}
\end{enumerate}

Each cluster's energy deposited $E$ and time of formation $t$ are calculated using calibrated attenuation length factors. The AI-assisted ECal clusters are passed back into \coatjava to yield a list of particles for the event.

\section{Results}
\label{sec:results}

To compare the clustering results of \coatjava and object condensation, we define a metric called "trustworthiness" for each reconstructed neutron. A neutron is considered trustworthy if it satisfies the following criteria:
\begin{enumerate}
    \item There is a true generated neutron within $\Delta\theta\leq 4^\circ$ and $\Delta\phi\leq 4^\circ$.
    \item There is no other reconstructed neutron within $\Delta\theta\leq 4^\circ$ and $\Delta\phi\leq 4^\circ$.
\end{enumerate}

In an ideal scenario, all reconstructed neutrons would meet the trustworthiness criteria. The first criterion validates that the reconstructed neutron is matched to a nearby generated neutron. Failing this condition could imply that the neutron was misidentified, or that the reconstructed ECal cluster attributed to the neutron was created by secondaries of other particles. The second criterion eliminates potential experimental ambiguity by confirming the absence of other candidate reconstructed neutrons in close proximity. In other words, the trustworthiness is the likelihood that a reconstructed neutron \textit{uniquely matches} to a nearby true neutron. A very low trustworthiness score indicates an overabundance of unreliable and experimentally difficult-to-validate neutrons. 

\begin{figure}[htbp]
    \centering
    \includegraphics[width=\textwidth]{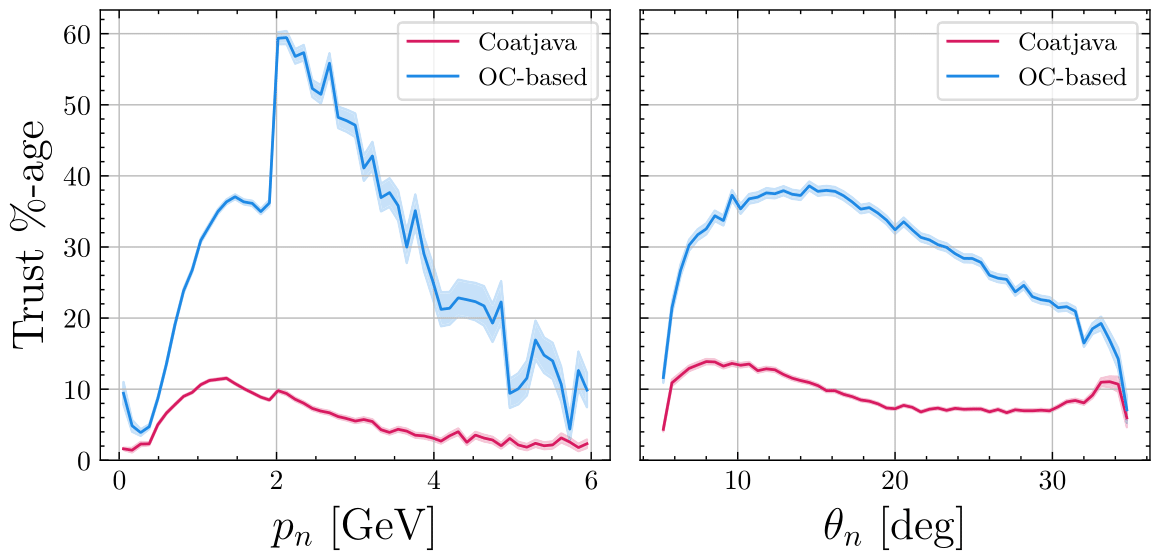}
    \caption{Trustworthiness of reconstructed neutrons as a function of momentum (left) and scattering angle (right). Error bands reflect uncertainty due to the number of Monte Carlo events sampled.}
    \label{fig:trust_2112}
\end{figure}

The trustworthiness of reconstructed neutrons was evaluated on the same simulated $e^-+p$ collision dataset described in Section \ref{sec:data}. The results are shown in Figure \ref{fig:trust_2112} binned in neutron momentum and neutron scattering angle. In total, $224,247$ distinct Monte Carlo neutrons left hits in the ECal across 1 million events. In the forward detector, base \coatjava reconstructs $858,984$ neutrons, of which $76,313$ ($8.88$\%) are trustworthy. Object condensation reconstructs $285,148$ neutrons, of which $87,631$ ($30.73$\%) are trustworthy. The AI-assisted clustering method developed more than triples the trustworthiness of reconstructed neutrons and greatly reduces the false neutron background.

A sudden change in the neutron trustworthiness can be seen just before $p=2$ GeV. This is an artifact of how neutrons are identified from a cluster later on in the reconstruction pipeline. When a neutral cluster (no matching track) is processed by \coatjava to determine if it is a neutron or photon, two criteria are checked:

\begin{enumerate}
    \item Is $\beta=v/c<0.9$? If so, particle is identified as a neutron.
    \item Is $\beta=v/c\geq0.9$ and this cluster contains no 3-way intersection in the PCal? If so, particle is identified as a neutron.
    \item Otherwise, particle is identified as a photon.
\end{enumerate}

The velocity $v$ of the particle is obtained using timing information from the event trigger and the calorimeter hits. This pre-existing classification algorithm provided by the collaboration thus groups neutrons into two categories: those with $v/c<0.9$ and those with $v/c>0.9$. At the neutron mass, $v/c=0.9$ corresponds to a momentum around $1.94$ GeV. Under close inspection, the functional behavior of OC-based trustworthiness (and even \coatjava) indeed changes around this value. Reconstructed neutrons without a PCal cluster may in general be more trustworthy as true neutrons tend to interact in the deeper layers of the ECal.

Furthermore, a steady decline in neutron-trustworthiness is observed for large momentum. At higher momentum, secondaries created by the hadronic interactions carry more energy, and thus produce more hits. This produces a higher risk of either clustering method inferring multiple distinct clusters, lowering trustworthiness. As for the $\theta$ dependence, dips in trustworthiness near $\theta_n=5^\circ,35^\circ$ can be attributed to the fact that true hadronic showers tend to "spill-out" near the detector edge, making an accurately reconstructed neutron less common.

\begin{figure}[htbp]
    \centering
    \includegraphics[width=\textwidth]{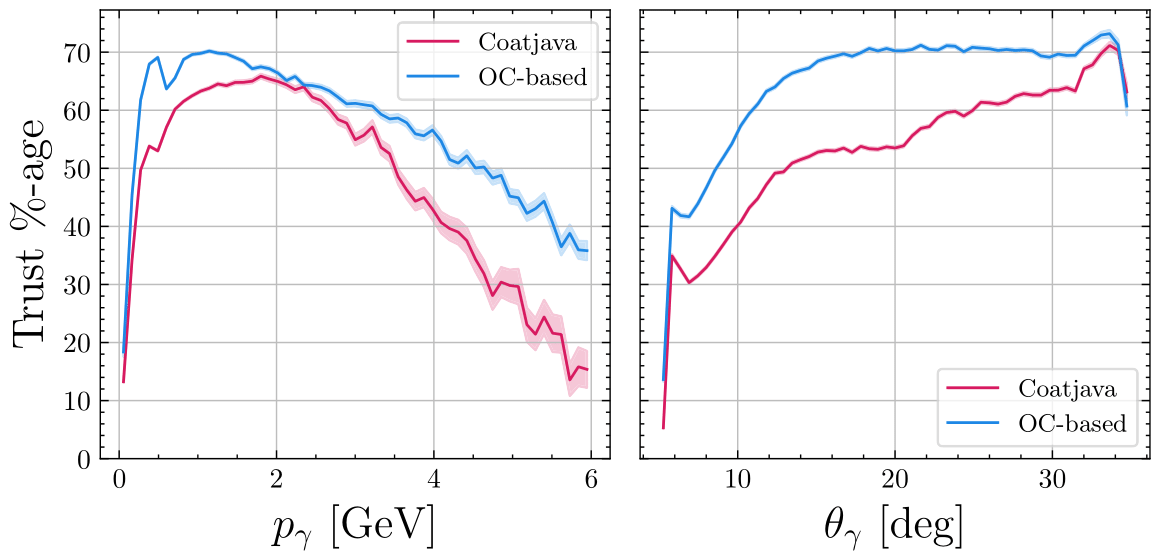}
    \caption{Trustworthiness of reconstructed photons as a function of momentum (left) and scattering angle (right). Error bands reflect uncertainty due to the number of Monte Carlo events sampled.}
    \label{fig:trust_22}
\end{figure}

In addition, the trustworthiness of reconstructed photons is shown in Figure \ref{fig:trust_22} binned in photon momentum and photon scattering angle. In total, $1,075,018$ distinct Monte Carlo photons left hits in the ECal. Of the $1,788,030$ forward detector photons reconstructed by base \coatjava, $913,128$ ($51.07$\%) are trustworthy. Object condensation reconstructs $1,358,349$ photons, of which $879,261$ ($64.73$\%) are trustworthy. A similar decrease in trustworthiness can be seen at higher momentum as the photon creates more secondaries, causing the algorithms difficulty in finding only one cluster. Interestingly, at large scattering angles, the photon trustworthiness becomes optimal. At large scattering angles, hadrons and photons are less common, so the resulting photon showers are more isolated and less likely to be entangled with the hits of other particles, making them easier to reconstruct accurately.

While object condensation does improve the trustworthiness of photons by $+13\%$, the overall number of trustworthy photons is about $4\%$ less. This decrease is related to a limitation of the object condensation model: its inability to assign a single detector hit to more than one particle cluster. Photons leave significantly fewer total hits in the ECal compared to hadrons like pions and neutrons. Thus, if the photon's sparse hits overlap with hits populated by hadrons in the same calorimeter sector, the photon's contribution can become "over-shadowed". This often results in a critical photon hit being incorrectly clustered with the hadron, preventing the formation of a distinct, three-way intersection of photon hits that would normally define the photon's cluster. \coatjava overcomes this by duplicating a strip object when it detects multiple intersections for it, thereby allowing these copies to be assigned to separate clusters. Recent advancements in computer vision --- particularly multi-object classification models like Mask R-CNN \cite{He_Gkioxari_Dollár_Girshick_2018} MaskFormer \cite{Cheng_Schwing_Kirillov_2021}, YOLACT \cite{Bolya_Zhou_Xiao_Lee_2019} and CondInst \cite{Tian_Shen_Chen_2020} --- offer promising strategies to overcome these limitations.

\section{Conclusion}
\label{sec:summary}

This paper presents a novel AI approach to ECal hit clustering at CLAS12 using object condensation. Our model uses both GravNet layers for local message-passing and a Transformer encoder for long-range message-passing. The trained network was integrated into the existing CLAS12 reconstruction pipeline, where it was shown to outperform previous methods in producing reliable neutron and photon clusters by a \textit{trustworthiness} metric. This improved trustworthiness can relax the need for overly conservative selection cuts in a range of exclusive measurements and opens the door for previously inaccessible neutron semi-inclusive DIS measurements.

To our knowledge, this study represents the first application of an AI-based clustering method to hodoscopic detectors. Our successful implementation widens the scope of clustering tasks that can be solved using AI. To improve the model, future work will be dedicated to exploring multi-object classification methods. We will also explore combining the clustering model with regression tasks such as reconstructing the cluster energy and position. The python project is available to view on GitLab \cite{comet}.

\section*{CRediT authorship contribution statement}
\textbf{Gregory Matousek:} Writing --- original draft, Investigation, Formal analysis, Data curation, Funding acquisition, Methodology, Software, Visualization. \textbf{Anselm Vossen:} Writing --- review \& editing, Supervision, Funding acquisition, Project administration.

\section*{Acknowledgments}
This material is based upon work supported by the U.S. Department of Energy, Office of Science, Office of Nuclear Physics under contracts DE-SC0024505 and DE-AC05-06OR23177. This research was supported by the National Science Foundation (NSF) under grant agreement DGE-2139754. We extend our gratitude to the DOE and NSF for their financial support, which made this work possible. We also thank the CLAS collaboration for providing the reconstruction framework and simulation environment for this project.


\end{document}